\documentclass[twocolumn,showpacs,preprintnumbers,amsmath,amssymb]{revtex4}

\usepackage{graphicx}
\usepackage{dcolumn}
\usepackage{bm}

\newcommand{\bq}{\begin{equation}}
\newcommand{\eq}{\end{equation}}
\newcommand{\bqa}{\begin{eqnarray}}
\newcommand{\eqa}{\end{eqnarray}}
\newcommand{\nn}{\nonumber \\}

\def\be     {\begin{equation}}
\def\ee     {\end{equation}}
\def\bea        {\begin{eqnarray}}
\def\eea        {\end{eqnarray}}
\def\bnn    {\begin{eqnarray*}}
\def\enn    {\end{eqnarray*}}

\begin{document}

\title{Role of non-magnetic disorder on the stability of $U(1)$ spin liquid : A renormalization group study}
\author{Ki-Seok Kim}
\affiliation{Korea Institute of Advanced Study, Seoul 130-012,
Korea}
\date{\today}

\begin{abstract}
Recently Hermele et. al claimed that the infrared (IR) fixed point
of non-compact $QED_3$ is stable against instanton excitations in
the limit of large flavors of massless Dirac fermions
[cond-mat/0404751]. We investigate an effect of non-magnetic
disorder on the deconfined quantum critical phase dubbed $U(1)$
spin liquid ($U1SL$) in the context of quantum antiferromagnet. In
the case of weak disorder the IR fixed point remains stable
against the presence of both the instanton excitations and
non-magnetic disorder and thus the $U1SL$ is sustained. In the
case of strong disorder the IR fixed point becomes unstable
against the disorder and the Anderson localization is expected to
occur. We argue that in this case deconfinement of spinons does
not occur since the Dirac fermion becomes massive owing to the
localization.
\end{abstract}


\maketitle

Recently  Hermele et. al pointed out that usual RPA treatment of
gauge fluctuations\cite{Permanent_confinement} is not sufficient
in order to examine an instanton effect even in the presence of
large flavors ($N$) of massless Dirac fermions in compact
$QED_3$\cite{Large_N_limit}. They showed that at the infrared (IR)
stable fixed point of non-compact spinor $QED_3$ in large $N$
limit instanton excitations become irrelevant and instanton
fugacity goes to zero. This originates from the fact that a
magnetic charge has a large value proportional to $N$ at the fixed
point. The large fixed point value of a magnetic charge is due to
screening of an electric charge by particle-hole excitations of
the massless Dirac fermions\cite{EM_duality}. The magnetic charge
goes to zero in the absence of the massless Dirac fermions at low
energy\cite{Large_N_limit}. The larger the magnetic charge, the
smaller the probability of instanton excitations. As a consequence
they concluded that deconfinement does exist at least at the
critical point in the large $N$ limit. In the context of quantum
antiferromagnet stable $U(1)$ spin liquid ($U1SL$) is
obtained\cite{Large_N_limit}.

In realistic cases disorder always exists. In the case of
non-interacting fermions it was shown by scaling argument that in
$3$ spatial dimension the presence of disorder causes a
metal-insulator transition\cite{Scaling}. But in $1$ or $2$
spatial dimension even weak disorder leads electrons to be
localized and only insulating phase is expected to
exist\cite{Scaling}. The presence of long range interaction can
change the above picture of non-interacting particles. Herbut
studied role of a random potential resulting from non-magnetic
disorder on a critical field theory of interacting bosons via
Coulomb interaction\cite{Herbut}. In the study he showed that
competition between the random potential and Coulomb interaction
leads to a new charged critical point near $3$ spatial dimension
where a dynamical critical exponent $z$ is exactly given by
$1$\cite{Herbut}. In the absence of the random potential the
charged critical point is not expected to exist and only the
standard runaway characteristic is found\cite{Herbut}. In the case
of fermi fields Ye investigated role of disorder on a Chern Simons
field theory of interacting Dirac fermions via Coulomb interaction
in $2$ spatial dimension\cite{Ye}. In the study he found a line of
fixed points which is stable in some cases\cite{Ye}. In the
absence of both the randomness and Chern Simons interaction only
the runaway characteristic was found as the case of bosons.

In this paper we investigate the role of non-magnetic disorder on
the deconfined quantum critical phase of $QED_3$ in the limit of
large flavors of massless Dirac fermions. In the concrete we
examine the stability of the IR fixed point of non-compact $QED_3$
in the presence of both the non-magnetic disorder and instanton
excitations. Existence of the stable IR fixed point in the absence
of disorder is a main difference from previous
works\cite{Herbut,Ye}. In the previous studies\cite{Herbut,Ye}
there are no stable IR fixed points in the absence of disorder as
discussed above. In the case of weak disorder the IR fixed point
is found to remain stable and $U(1)$ spin liquid ($U1SL$) in the
context of quantum antiferromagnet is expected to survive. The
stability against the weak disorder results from the existence of
the IR fixed point in the absence of disorder in $3$ space and
time dimension. In the case of strong disorder we find that it
becomes unstable. We are led to strong coupling regime where the
Anderson localization is expected to occur. Owing to the
localization the Dirac fermions become massive. In the case of
massive Dirac fermions usual RPA treatment may be possible.
Instantons are expected to be proliferated. We argue that
deconfinement of spinons is not expected to exist in the strong
disorder. In addition, we discuss a bosonic field theory in the
presence of non-magnetic disorder and find a difference in role of
disorder on the Dirac fermions and bosons respectively.

First we review deconfinement at the IR fixed point of $QED_3$ in
the absence of non-magnetic disorder\cite{Large_N_limit}. We
consider an effective action usually called $QED_3$ in imaginary
time \bqa && S = \int{d^Dx} \Bigl[\sum_{\sigma = 1}^{N}
\bar{\psi}_{\sigma}\gamma_{\mu}(\partial_{\mu} -
ia_{\mu})\psi_{\sigma} + \frac{1}{2e^2}|\partial\times{a}|^2 .
\Bigr]\eqa Here $\psi_{\sigma}$ is a massless Dirac spinor with a
flavor index $\sigma = 1, ..., N$ and $a_{\mu}$, a compact U(1)
gauge field. $x = ({\bf r},\tau)$ with $D-1$ dimensional space
${\bf r}$ and imaginary time $\tau$. In the context of U(1) slave
boson representation of SU(N) quantum antiferromagnet this action
can be considered as an effective action in the $\pi$ flux
phase\cite{Don_Kim}. In this case the Dirac spinor represents a
spinon carrying only the spin quantum number $1/2$. It is well
known that $QED_3$ with non-compact U(1) gauge field has a stable
IR fixed point in the limit of large flavors of massless Dirac
fermions\cite{KT}. A renormalization group (RG) equation for an
electric charge is easily obtained to be in one loop
order\cite{KT,Large_N_limit}, \bqa &&\frac{de^2}{dl} = (4-D)e^2 -
\lambda{N}e^4 , \eqa where $\lambda$ is a positive numerical
constant. The first term represents a bare scaling dimension of
$e^2$. In $(2+1)D$ $e^2$ is relevant in contrast with $(3+1)D$
where it is marginal. The second term originates from self-energy
correction of the U(1) gauge field by particle-hole excitations of
massless Dirac fermions. As shown by this RG equation, a stable IR
fixed point of $e^{*2} = \frac{1}{\lambda{N}}$ exists in $QED_3$.
Now our question is if the IR fixed point remains stable after
admitting instanton excitations. Using the electromagnetic
duality, Hermele et. al obtained RG equations of a magnetic charge
$g = \frac{1}{e^2}$ and an instanton fugacity $y_m$, \bqa &&
\frac{dg}{dl} = -(4-D)g - \alpha{y_m^2}{g^3} + \lambda{N} , \nn &&
\frac{dy_m}{dl} = (D - \beta{g})y_{m}, \eqa where $\alpha$ and
$\beta$ are positive numerical constants\cite{Large_N_limit}.
Owing to the last term $\lambda{N}$ in the first equation a
magnetic charge has a large fixed point value proportional to $N$,
i.e., $g^* = \lambda{N}$. As a consequence the instanton fugacity
goes to zero at this IR fixed point. $U1SL$ in terms of a spinon
(Dirac fermion) and non-compact U(1) gauge field is obtained at
the quantum critical phase.

Next we investigate the stability of the $U1SL$ fixed point in the
presence of non-magnetic disorder. We reconsider $QED_3$ in the
presence of non-magnetic disorder \bqa && S = \int{d^Dx}
\Bigl[\sum_{\sigma = 1}^{N}
\bar{\psi}_{\sigma}\gamma_{\mu}(\partial_{\mu} -
iea_{\mu})\psi_{\sigma} + \frac{1}{2}|\partial\times{a}|^2 \nn &&
+ V(x)\sum_{\sigma =
1}^{N}\bar{\psi}_{\sigma}\gamma_{0}\psi_{\sigma} \Bigr] . \eqa
Here $V(x)$ is a random potential generated by non-magnetic
disorder. It couples to a spinon density owing to the relation of
$V\sum_{\sigma=1}^{N}c_{\sigma}^{\dagger}c_{\sigma} =
V\sum_{\sigma=1}^{N}f_{\sigma}^{\dagger}f_{\sigma}$\cite{Don_Kim}.
Here $c_\sigma$ represents an electron with spin $\sigma$ and
$f_\sigma$, a spinon with spin $\sigma$. A physically relevant
case is that the random potential is random only in space but
static in time. Thus it does not depend on imaginary time, i.e.,
$V(x) = V({\bf r})$. We assume that $V({\bf r})$ is a gaussian
random potential with $<V({\bf r})V({\bf r'})> = W\delta({\bf r} -
{\bf r'})$ and $<V({\bf r})> = 0$\cite{Herbut,Ye}. Using the
standard replica trick to average over the gaussian random
potential, we obtain an effective action in the presence of
non-magnetic disorder \bqa && S = \int{d^{D-1}{\bf r}}{d\tau}
\Bigl[\sum_{\alpha=1}^{M}\Bigl(\sum_{\sigma = 1}^{N}
\bar{\psi}_{\sigma,\alpha}\gamma_{\mu}(\partial_{\mu} -
iea_{\mu,\alpha})\psi_{\sigma,\alpha} \nn && +
\frac{1}{2}|\partial\times{a}_{\alpha}|^2 \Bigr)\Bigr] \nn && -
\frac{W}{2}\sum_{\alpha,\alpha' = 1}^{M}\sum_{\sigma,\sigma' =
1}^{N}\int{d^{D-1}{\bf r}}{d\tau_1}{d\tau_2} \nn &&
\bar{\psi}_{\sigma,\alpha}({\bf
r},\tau_1)\gamma_{0}\psi_{\sigma,\alpha}({\bf
r},\tau_1)\bar{\psi}_{\sigma',\alpha'}({\bf r},\tau_2
)\gamma_{0}\psi_{\sigma',\alpha'}({\bf r},\tau_2) . \eqa Here
$\alpha, \alpha'$ is replica indices and the limit $M \rightarrow
0$ is to be taken at the end.

Introducing renormalized field variables of $\psi_{\sigma} =
e^{-\frac{D+z-2}{2}l}Z_{k}^{1/2}\psi_{\sigma,r}$ and $a_{\mu} =
e^{-\frac{D+z-3}{2}l}Z_{a}^{1/2}a_{\mu,r}$, we obtain renormalized
couplings of $e^{2} = e^{-(5-D-z)l}Z_{a}^{-1}e_{r}^2$ and $W =
e^{-(4 - D - z)l}Z_{k}^{-2}Z_{W}W_{r}$. Here $z$ is a dynamical
critical exponent. $Z_k, Z_a, Z_W$ are usual renormalization
constants of a Dirac fermion, gauge field and strength of a random
potential respectively. A subscript $r$ represents "renormalized".
Eq. (5) is obtained to be in terms of renormalized variables \bqa
&& S = \int{d^{D-1}{\bf r}'{d\tau'}}
\Bigl[\sum_{\alpha=1}^{M}\Bigl(\sum_{\sigma = 1}^{N}
Z_{k}\bar{\psi}_{\sigma,\alpha}\gamma_{\mu}(\partial_{\mu}' -
iea_{\mu,\alpha})\psi_{\sigma,\alpha} \nn && +
\frac{Z_{a}}{2}|\partial{'}\times{a_{\alpha}}|^2 \Bigr)\Bigr] -
Z_{W}\frac{W}{2}\sum_{\alpha,\alpha'=1}^{M}\sum_{\sigma,\sigma' =
1}^{N}\int{d^{D-1}{\bf r}'}{d\tau_{1}'}{d\tau_{2}'} \nn &&
\bar{\psi}_{\sigma,\alpha}({\bf
r}',\tau_{1}')\gamma_{0}\psi_{\sigma,\alpha}({\bf
r}',\tau_{1}')\bar{\psi}_{\sigma',\alpha'}({\bf r}',\tau_{2}'
)\gamma_{0}\psi_{\sigma',\alpha'}({\bf r}',\tau_{2}') \eqa with
rescaled space ${\bf r}' = e^{-l}{\bf r}$ and time $\tau' =
e^{-zl}\tau$. In the above we omitted a subscript $r$ for a simple
notation. Calculating the renormalization constants $Z_k, Z_a,
Z_W$ in one loop order, we obtain RG equations \bqa
&&\frac{de^2}{dl} = (5 - z - D)e^2 - \lambda{N}e^4 , \nn &&
\frac{dW}{dl} = (4 - z - D - \chi{e}^{2})W + \zeta(N+c){W}^{2}
\eqa with positive numerical constants $\lambda, \chi, \zeta, c$.
Here $z$ is determined by the condition of $e^{-2zl}Z_{\omega} =
e^{-2l}Z_{k}$ with $Z_k$, a renormalization constant of a Dirac
fermion in momentum and $Z_w$, that in energy\cite{Herbut}, which
gives \bqa && z = 1 + AW \eqa with a positive numerical constant
$A$. Our interest is the case of $D = 2 + 1$, i.e., $2$ space and
$1$ time dimension. These RG equations basically coincide with
those of Ref. [6] in $D = 2 + 1$ if the term $(4 - D)e^2$ in the
first RG equation is neglected. The presence of this term leads to
the stable IR fixed point as discussed earlier. Precise values of
the positive numerical constants are not important in our
consideration. We solve the above RG equations with arbitrary
positive numerical constants in order to understand a general
structure.

First we check the case of non-interacting Dirac fermions in the
presence of a random potential. In zero charge limit ($e
\rightarrow 0$) a RG equation is given by \bqa \frac{dW}{dl} = (4
- z - D )W + \zeta(N+c){W}^{2} . \eqa In $D = 3 + 1$ ($3$ spatial
dimension) there is an unstable fixed point of $W_c =
\frac{1}{\zeta(N+c)}$ with $z = 1 + \frac{A}{\zeta(N+c)}$ to
separate a metal and an insulator. In $D = 2 + 1$ of present
interest the unstable fixed point becomes zero, i.e., $W_c = 0$.
Thus only insulating phase is expected to exist\cite{Scaling} as
discussed in the introduction.

In the case of interacting Dirac fermions via long range
"electromagnetic" interaction\cite{Interaction} these RG equations
[Eq. (7)] show three fixed points in $D = 2 + 1$; the first is
$W_{1c} = 0$ and $e_{1c}^{2} = 0$ with $z=1$, the second, $W_{2c}
= 0$ and $e^{2}_{2c} = \frac{1}{\lambda{N}}$ with $z=1$, and the
third, $W_{3c} = O(\frac{1}{N^2})$ and $e_{3c}^{2}
=\frac{1}{\lambda{N}} + O(\frac{1}{N^3})$ with $z = 1 +
O(\frac{1}{N^2})$. The first is a fixed point of free Dirac
fermions which is unstable for non-zero charge
$e^2$\cite{Dirac_fixed_point}. The RG flow goes to the second, the
IR fixed point of non-compact $QED_3$. The third is an unstable
fixed point. This fixed point does not exist in the absence of the
gauge interaction. Existence of the new unstable fixed point
originates from the term $(4-D)e^2$ of the first RG equation in
Eq. (7), representing relevance of an electric charge in $D =
2+1$. In the case of small strength of the random potential, i.e.,
$W < W_{3c}$, the random potential becomes irrelevant and the
usual IR fixed point (the second) remains stable. The stability
against the weak disorder results from the existence of the stable
IR fixed point in the absence of disorder in $D = 2+ 1$, as will
be shown below. In the case of large strength of the random
potential, i.e., $W > W_{3c}$, the random potential becomes
stronger. We are led to strong coupling regime where our
perturbative calculation does not apply. In this case the Anderson
localization is expected to occur and thus the Dirac fermions
become massive.

In order to see the stability of the fixed points we expand the RG
equations near each fixed point. Inserting $e^2 = e_{2c}^{2} + f$
and $W = W_{2c} + h$ to Eq. (7), we obtain linearized RG equations
near the IR fixed point \bqa && \frac{df}{dl} = -f -
\frac{A}{\lambda{N}}h , \nn && \frac{dh}{dl} = -
\frac{\chi}{\lambda{N}}h . \eqa As shown by these RG equations, it
is clear that the IR fixed point is stable against the weak
disorder. The stability originates from the finite fixed point
value of an electric charge, i.e., $e_{2c}^{2} =
\frac{1}{\lambda{N}}$. Expanding the RG equations near the third
fixed point, we obtain linearized RG equations to the order of
$\frac{1}{N}$ \bqa && \frac{df}{dl} = -f - \frac{A}{\lambda{N}}h ,
\nn && \frac{dh}{dl} = \frac{\chi}{\lambda{N}}h . \eqa The second
equation shows the instability of this fixed point.

Now we examine an instanton effect on these fixed points. Using
the electromagnetic duality, we obtain RG equations of a magnetic
charge $g = e^{-2}$ and an instanton fugacity $y_{m}$ in the
presence of a random potential \bqa && \frac{dg}{dl} = -(5 - z -
D)g - \alpha{y_m^2}{g^3} + \lambda{N} , \nn && \frac{dW}{dl} = (4
- z - D - \chi\frac{1}{g})W + \zeta(N+c){W}^{2} , \nn &&
\frac{dy_m}{dl} = (z - 1 + D - \beta{g})y_{m} \eqa with positive
numerical constants $\alpha,\beta$ same as those in Eq. (3). The
term $- \alpha{y_m^2}{g^3}$ in the first RG equation is added
owing to screening of a magnetic charge by instanton
excitations\cite{Large_N_limit}. In the case of weak disorder,
i.e., $W < W_{3c}$, the IR fixed point of $g^{*} = \lambda{N}$
remains stable as discussed above. Thus the $U1SL$ is sustained by
the same reason as the case of no disorder. In the case of strong
disorder, i.e., $W > W_{3c}$, the IR fixed point becomes unstable.
The strength of disorder becomes larger. An important question is
whether the magnetic charge goes to zero or not as the strength of
disorder gets larger. At first glance of Eq. (11) the fixed point
value of an electric charge seems to be sustained. Thus one may
conclude that $U1SL$ is still expected to occur. In this case,
more precisely, gapped $U1SL$ may appear owing to the Anderson
localization which is expected to occur in the strong disorder.
But this is an illusion. As discussed above, the Dirac fermions
are expected to be massive owing to the localization. The above
calculation cannot apply to this strong coupling regime. Further,
this phase is not expected to be critical any more. In this
"insulating" phase of spinons screening of internal charge becomes
negligible. The internal charge is expected to go to infinity
following its bare scaling dimension. The magnetic charge goes to
zero. In this case instanton excitations are relevant perturbation
and the instanton fugacity gets larger to go to infinity. Thus
deconfinement of spinons is not expected to occur in the strong
disorder.

Now we discuss an effect of non-magnetic disorder on a critical
field theory of interacting bosons via long range electromagnetic
interaction. The critical field theory usually dubbed scalar
$QED_3$ can be considered to describe interacting holons via
internal gauge interaction in the context of $U(1)$ slave boson
theory\cite{Don_Kim,Kim}. In the scalar $QED_3$ a stable IR fixed
point called charged or IXY fixed point is also found in the limit
of large $N$\cite{Kleinert} as the case of the spinor $QED_3$.
Here $N$ is the flavor number of boson fields. The charged fixed
point governs a superconductor to insulator transition. The
charged fixed point is shown to be unstable in the presence of
weak disorder\cite{Boson_disorder}. A new stable fixed point is
expected to appear in association with a random potential. This
fixed point seems to be related with a Bose glass to
superconductor transition\cite{Herbut,Boson_disorder}, not a Mott
insulator to superconductor transition governed by the charged
fixed point\cite{Kleinert}. But its nature is not clear. Emergence
of the new stable fixed point is a main difference between the
scalar and spinor $QED_3$. This is because even weak disorder is a
relevant perturbation only in a bosonic field theory. The
relevance of weak disorder originates from different scaling
between bosons and Dirac fermions\cite{Scaling_dimension}. This
different scaling is due to a difference in equation of motion,
the Klein-Gordon equation and Dirac equation respectively. At this
new stable fixed point the fixed point value of an electric charge
square is still proportional to inverse of the flavor number of
bosons, i.e., $e_{c}^{2} \sim N^{-1}$\cite{Boson_disorder}. Thus a
magnetic charge is proportional to $N$ as the case of the charged
fixed point. In the limit of large flavor number instanton
fugacity goes to zero at this new fixed point. Deconfinement of
boson fields (holons) is expected to occur. Bosonic U(1) liquid is
sustained at the new fixed point associated with non-magnetic
disorder.

In this paper we considered SU(N) quantum antiferromagnet
described by $N$ flavors of massless Dirac fermions interacting
via compact U(1) gauge fields. But a real antiferromagnet has
SU(2) symmetry. Thus the flavor number of the Dirac fermions is
given by $N = 2$\cite{Don_Kim}. In this case it is not clear
whether the present result can be applicable. It is known that
there exists a critical flavor number $N_c$ associated with
spontaneous chiral symmetry breaking ($S\chi{S}B$) in
$QED_3$\cite{Don_Kim}. But a precise value of the critical number
is still in debate\cite{N_c}. If the critical value is larger than
$2$, the $S\chi{S}B$ is expected to occur for the physical $N = 2$
case\cite{Don_Kim,Kim_CSB}. The Dirac fermions become massive. In
the $S\chi{S}B$ phase the massive Dirac fermions are confined to
form mesons, here, magnons\cite{Don_Kim}. As a result, in the case
of $N_c > N = 2$ the $U1SL$ is not expected to exist and the
present consideration is not applied. But there is a cure even in
this case. We consider hole doping to the $U1SL$. Doped holes are
represented by holons carrying only charge degree of
freedom\cite{Senthil_Lee}. Recently Senthil and Lee claimed that
critical fluctuations of holons at a quantum critical point
associated with a superconducting transition can result in
suppression of instanton excitations and thus the quantum critical
point can be described by the $U1SL$ for spin degree of
freedom\cite{Senthil_Lee}. This argument is based on the fact that
the $S\chi{S}B$ does not occur owing to critical fluctuations of
holons. The critical fluctuations increase a flavor number of
massless fluctuations\cite{N_c_holon}. If a total flavor number of
massless spinons and holons exceeds the critical value $N_c$, the
$S\chi{S}B$ is not expected to
occur\cite{N_c_holon,Critical_holon}. As a result the present
scenario for the role of non-magnetic disorder in the $U1SL$ has a
chance to be applicable.

To summarize, we showed that the $U(1)$ spin liquid is sustained
against the presence of weak non-magnetic disorder. However,
strong disorder leads the fixed point to be unstable and the RG
flow goes to strong coupling regime. In the strong coupling regime
our perturbative RG does not work any more. Thus more refined
calculation is required. We argued that deconfinement of spinons
does not occur in this regime since the Dirac fermions become
massive owing to localization. Lastly, we compared the case of
Dirac fermions with that of bosons. In a bosonic field theory a
new stable fixed point associated with disorder emerges in
contrast with a fermionic field theory. At this new stable fixed
point an electric charge is sufficiently screened and
deconfinement of bosons is expected to occur.

K.-S. Kim thanks prof. Han, Jung-Hoon for helpful discussions of
basic concept in association with disorder.


\begin{thebibliography}{9}
\bibitem{EM_duality} Consider the electromagnetic duality of $eg
= 1$ where $e$ is an electric charge and $g$, a magnetic charge.
If an electric charge becomes small owing to screening effect by
massless Dirac fermions, a magnetic charge gets large.
\bibitem{Interaction} Here the electric charge $e$ is related
with not a real electromagnetic field $A_{\mu}$ but an internal
U(1) gauge field $a_{\mu}$\cite{Don_Kim}.
\bibitem{Dirac_fixed_point} In the case of Coulomb
interaction the fixed point of free Dirac fermions is stable
against small charge\cite{Ye}. This is because there is no term
like $(4-D)e^2$ of the first RG equation in Eq. (7).
\bibitem{Scaling_dimension} From a bosonic action of $S =
\int{d^{D-1}r}{d\tau}|\partial_{\mu}\phi|^2$ a scaling is easily
determined by $\phi = e^{-\frac{D+z-3}{2}l}Z_{k}^{1/2}\phi_{r}$
with a renormalized field $\phi_r$ and a renormalization constant
$Z_k$. We note the difference between the Dirac fermion and boson
by the exponent $\frac{D+z-2}{2}$ and $\frac{D+z-3}{2}$
respectively. This difference changes $-\chi{e}^{2}W$ at $D = 3$
in Eq. (7) into $(2 - \chi{e}^{2})W$ in Eq. (6) of Ref. [5].
\bibitem{N_c_holon} A flavor number of massless fluctuations is
given by $N = N_f + N_b$. Here $N_f$ is the flavor number of
massless Dirac fermions and $N_b$, that of critical holon
fluctuations. In a SU(2) slave boson theory\cite{Senthil_Lee}
$N_b$ is given by $2$. Thus the total flavor number is $N = 4$.
This exceeds a known critical value $N_c \approx
3.24$\cite{Don_Kim}. Chiral symmetry breaking is not expected to
occur.
\end{thebibliography}
\end{document}